\documentstyle[aps,prl,twocolumn,epsf,rotate,bookmath]{revtex}
\begin{document}

\draft
\twocolumn[\hsize\textwidth\columnwidth\hsize\csname @twocolumnfalse\endcsname

\title{
Time Reversal Symmetry Breaking States Near Grain Boundaries \\
Between d-wave Superconductors 
} 
\author{M. Fogelstr\"om and S.-K. Yip}
\address{ 
Department of Physics \& Astronomy, 
           Northwestern University, Evanston, Illinois 60208, U. S. A. \\
Department of Physics, {\AA}bo Akademi, 
           Porthansgatan 3, 20500 {\AA}bo, Finland} 
\date{\today}

\maketitle

\begin{abstract}

{In this paper we study the order parameter and density
of states near a grain boundary between two $d_{x^2-y^2}$
superconductors.  We examine  
broken time reversal symmetry near the interface.
In particular we show that,
under suitable circumstances, time reversal symmetry
must be broken even when the order parameter
is purely $d_{x^2-y^2}$ everywhere in space.

 }
\end{abstract}
] %End of two-column style title mode

The $d_{x^2-y^2}$ order parameter, appropriate to the hole-doped oxide superconductors,
preserves time-reversal symmetry (TRS) in the bulk. 
At surfaces and  interfaces it is now known that time-reversal symmetry
may be broken.
NIS-tunneling experiments by Covington {\it et~al}. 
\cite{Covington97} indicate a time reversal symmetry breaking
(TRSB) state locally at surfaces.\cite{Fogelstrom97a}
Fractional fluxes at corners of interfaces in inclusion experiments by 
Kirtley {\it et~al.} \cite{Kirtley96} strongly indicate
that TRS may also be broken at grain boundaries.
\cite{Sigrist95,Yip95,Sigrist96,Bailey97}
In this paper we discuss the origin of the 
TRSB-state and constrast TRSB at
low and high transmission interfaces.

At a surface
or interface with low transmission,
TRSB can occur in the presence
of subdominant pairing interaction in channels
other than the dominant $d_{x^2-y^2}$. In this 
case order parameters corresponding
to those channels can appear near the interface 
\cite{Buchholtz95,Matsumoto95}.  
For example, if a subdominant pairing interaction is
present in the s-wave channel, then, under suitable conditions,
the order parameter near the interface can have a
$d \pm i s$ symmetry.
The order parameter thus  breaks TRS locally. 
  A prerequisite for the TRSB-state
is substantial pair-breaking at the interface, 
{\it i.e.} the misorientation of the
surface normal to crystal a-axis should be close to 
$\frac{\pi}{4} \ ({\rm mod} \ \pi/2)$.

The above is in constrast to
the case where there is a reasonably high transmission 
probability of electrons across the interface.
In this case the subdominant pairing interaction is not necessary
for TRSB at the interface.\cite{Yip95,Barash95,Fogelstrom97b}
TRSB occurs even when the order parameter is purely $d_{x^2-y^2}$ everywhere
near the interface. The origin of this TRSB-state is a proximity effect and
it arises because the minimum
energy state for the interface corresponds to a state
with a finite phase difference, $\Delta\chi=\chi_R - \chi_L$,
across the junction.  Here $\chi_L$ and $\chi_R$ are the phases of the order parameter
on either side far away from the interface. $\Delta \chi$ is
other than an integral multiple of $\pi$ for the TRSB-state.  
In this case states with minimum total
interface free energy occur in
pairs related by time-reversal: if $\Delta \chi$ corresponds to
a state with  minimum  energy, there is also a non-equivalent but
degenerate state with $ \Delta \chi' = - \Delta \chi$.
An interface at its minimum energy configuration 
will have TRS spontaneously broken. 
In contrast to the one discussed in the last paragraph, this 
is the more likely route to TRSB when the
transmission probability across the interface is moderate to high
and when the misorientation between the two superconductors
are close to $\pi/4 \ ({\rm mod} \ \pi/2)$.
\cite{Fogelstrom97b,Yip97} 

As is well-known, Josephson effects occur 
in the presence of an interface with
finite transmission.
At a general phase difference between the two superconductors,
a dissipationless current, $J_x$, can flow through the interface.
If the order parameter
itself does not break TRS, then the current across 
the interface is always zero for $\Delta \chi$  
being an integral multiple of $\pi$. However,  
under appropriate conditions there can be additional
values of $\Delta\chi$ where the net current
across the interface vanishes.  Previously one of us
\cite{Yip95} has explained
how this can occur for a pinhole junction by considering
the sum of contributions to the current from different
parts of the Fermi surface.  These new states with zero 
net current 
occur as a combined result of non-sinusoidal current-phase relationships
and sign changes of the order parameter. 
The $J_x = 0$ states correspond to states where the 
 junction energy is at a relative extrema
as a function of the phase difference $\Delta \chi$.
In particular it can be shown that the new $J_x = 0$ states,
if they exist, correspond to energy minima. \cite{Yip95}

In this paper we study  a planar interface with uniform transmission.
In general a current flows through the interface.
The corresponding states possess finite flow energy densities even
far away from the interface.  Here we focus on
the set of states with $ J_x = 0$ for which this contribution
is absent. In these states
the gradient of the phase of the d-wave order parameter
vanishes as $ x \to \pm \infty$.  
We shall show that many of the statements concerning the new
energy miminum states mentioned 
above for the pinhole\cite{Yip95}
are still correct for the planar interface, provided
appropriate minor modifications are made. 
If the order parameter is purely $d_{x^2-y^2}$,
it is easy to verify that states which correspond to $J_x = 0$
with $\Delta \chi = 0$
or $\pi$ with  $\chi$ piecewise constant
are always possible. 
At not too small transmission
across the interface, there may be other states with different
$\Delta \chi$ which also correspond to $J_x = 0$ and  
under appropriate conditions
states with $\Delta \chi \ne 0 $ or $\pi$ will 
correspond to the minimum energy.
We shall compare the free energies, order parameters and 
densities of states (DOS) of these $J_x = 0$  states.
Apart from its intrinsic interest, we
shall see that the DOS 
provides an alternative view of the mechanisms for TRSB.
A signature of a TRSB-state is that the ZEBS at $\varepsilon=0$
are shifted away from the midgap and that spontaneous currents along
the interface are nucleated.\cite{Buchholtz95,Matsumoto95}

The occurrence of  zero energy bound states (ZEBS)
for non-transmitting surfaces have already been extensively
investigated 
%\cite{Buchholtz81,Hu94,Tanaka95,
\cite{Fogelstrom97a} (and references therein).  
For order
parameters real up to a gauge transformation, 
ZEBS are present for
the quasiparticle paths along which a sign change of 
the order parameter occurs. 
ZEBS are also common for interfaces with finite
transmission if TRS is preserved
($\Delta \chi = 0$ or $\pi$). 
 One can show  rigorously \cite{proof} that
 ZEBS are present irrespective
of the value of the transmission coefficient
 whenever there are quasiparticle paths such that
a quasiparticle experiences a sign change of the order parameter
both if it is transmitted or reflected.
 For interfaces between superconductors with large
misorientation and in states which preserve TRS,
ZEBS occur over a large 
part of the Fermi surface.
The existence of these low energy bound states
corresponds to severe pair-breaking 
near the interface.
These ZEBS can be pushed to finite energies
by allowing a finite phase difference between the 
two superconductors.  
Correspondingly we shall show that the magnitude
of the order parameter for the TRSB state 
(denoted simply by $\Delta \chi \ne 0$ below) 
is larger than  the 
corresponding states with $\Delta \chi = 0$ or $\pi$.
The formation  of ZEBS and the suppression of
the order parameter near the interface
suggest that the $\Delta \chi = 0$ or $\pi$ states are
energetically unfavorable compared with the TRSB state.
 \cite{notebruder}
This is verified  by a calculation of
the free energy.

\begin{minipage}{0.48\textwidth}
%%%%%%%%%%%%%%%%%%%  Table 1. Junction energies %%%%%%%%%%%%%%%%%%%%%%%%%%%%%%%%%%%%%%%%%
\begin{table}[]
\caption[]{Free energy per unit surface area calculated as 
$\Delta {\cal F}={\cal F}_{junc}-{\cal F}_{bulk}$
for the junctions shown. The unit of $\Delta {\cal F}$ is 
$N_f (\hbar v_f) (2\pi T_c)$. $N_f$ is the normal state DOS.}
\begin{tabular}{ccccc}
$(\alpha_L,\alpha_R)$&State & $\Delta {\cal F}$& $\Delta {\cal F}$& $\Delta {\cal F}$\\ 
                     &      & ${\cal D}_o=1.0$ & ${\cal D}_o=0.7$ & ${\cal D}_o=0.3 $\\
\hline
\hline
$ (0,\frac{\pi}{4})$               &$\Delta\chi=0$  &0.127&0.125&0.116 \\
                                   &$\Delta\chi\ne0$&0.109&0.119&0.115 \\
%                                  &$d+is$          &0.107&0.115&0.113 \\%
\hline
$(-\frac{\pi}{12},\frac{\pi}{6})$  &$\Delta\chi=0$  &0.133&0.132&0.130 \\
                                   &$\Delta\chi=\pi$&0.120&0.119&0.123 \\
                                   &$\Delta\chi\ne0$&0.108&0.117&0.123 
                                   
\end{tabular}
\label{tab:energies}
\end{table}
%%%%%%%%%%%%%%%%%%%  Table 1. Junction energies %%%%%%%%%%%%%%%%%%%%%%%%%%%%%%%%%%%%%%%%%

\end{minipage}
\vspace*{0.05truecm}

For definiteness, we model the interface
as an ideal, smooth barrier with 
a delta function potential.
In this case, the interface can be parameterized by  ${\cal D}_o$,
the coefficient of transmission
for normal incidence.  The transmission coefficient
${\cal D}(\phi)$ for momenta $\hat \vp_f$
at an angle $\phi$ with respect
to the interface normal is given by
\be
{\cal D}(\phi) = { {\cal D}_o { \rm cos}^2 \phi \over
    1 -  {\cal D}_o { \rm sin}^2 \phi }. 
\label{transprop}
\ee
The quasiclassical green's function $\hat g$ obeys
the usual equation involving $\hat \Delta$
in the bulk, and boundary conditions at the interface
parameterized by ${\cal D}(\phi)$ (see Refs. \cite{Fogelstrom97b} and \cite{Yip97}
for details).  
The numerical procedure is as follows: We start with
an initial ansatz of the order parameter $\hat \Delta (x)$,
which in general possesses a phase difference
far away from the interface.
For the given $\hat \Delta (x)$, we obtain the quasiclassical
green's function
$\hat g (\hat \vp_f, \varepsilon, x)$ 
along the imaginary axis (here $\varepsilon = i \varepsilon_n$ where
$\varepsilon_n$'s are the Matasubara frequencies)
via the method described in
detail in \cite{Yip97}.  No iteration is needed in this step.
The correction to $\hat \Delta (x)$ is determined
by the weak-coupling gap equation involving
the off-diagonal parts of $\hat g$.  At each iteration
the current across the interface is also calculated.
Then a gauge transformation 
depending on the calculated current   
is performed on the order parameter to relax
$\hat \Delta$ towards the state with $J_x = 0$ 
with the desired sign of $ {\partial J_x \over \partial \Delta \chi} $.
This procedure is repeated until $\hat \Delta (x)$ is consistent
with the calculated $\hat g$ and the condition $J_x =0$.
Note that once self-consistency is achieved, particle
conservation will be respected (see, e.g., \cite{Serene83}).
$J_x $ is then $x$ independent and thus $ J_x = 0$ at all $x$.
After obtaining the self-consistent order parameter
we evaluate
the free energy via the "$\lambda $ trick". \cite{Thuneberg84}

With the self consistent order parameter the green's function
$\hat g$ was evaluated on the real energy axis 
again via the method in \cite{Yip97}.
From this $\hat g$ we obtain the density of states.
All the DOS  below are obtained at
energies $\varepsilon+i\gamma$, with $\gamma=0.05T_c$ . 
This small imaginary part of the energy  
simulates a broadening of energy levels that
would occur naturally in  non-ideal systems.

We first confine ourselves to pure $d_{x^2-y^2}$
order parameter. We write  
\be
\Delta (\hat \vp_f, x) =  \eta_d(x) {\sqrt 2} \cos (2 (\phi -\alpha)),
\label{purestate}
\ee
which defines the complex order parameter $ \eta_d (x)$
with $\alpha = \alpha_L$ or $\alpha_R$ for the left and
right side of the interface respectively. 
Here $\alpha_L$ and $\alpha_R$  denote the orientations
of the crystals on the two sides of the interface.
They specify
the angle between the $\hat a$ axis and hence 
the positive lobe of the order parameter with respect
to the normal to the interface.
We find that
TRSB is most significant at low temperatures and when the misorientation
 $\theta = \alpha_R - \alpha_L$
is close to $\pi/4 \ ({\rm mod} \ \pi/2$).

\begin{figure}[h]
\centerline{\epsfysize=.48\textwidth \rotate[r]
{\epsfbox{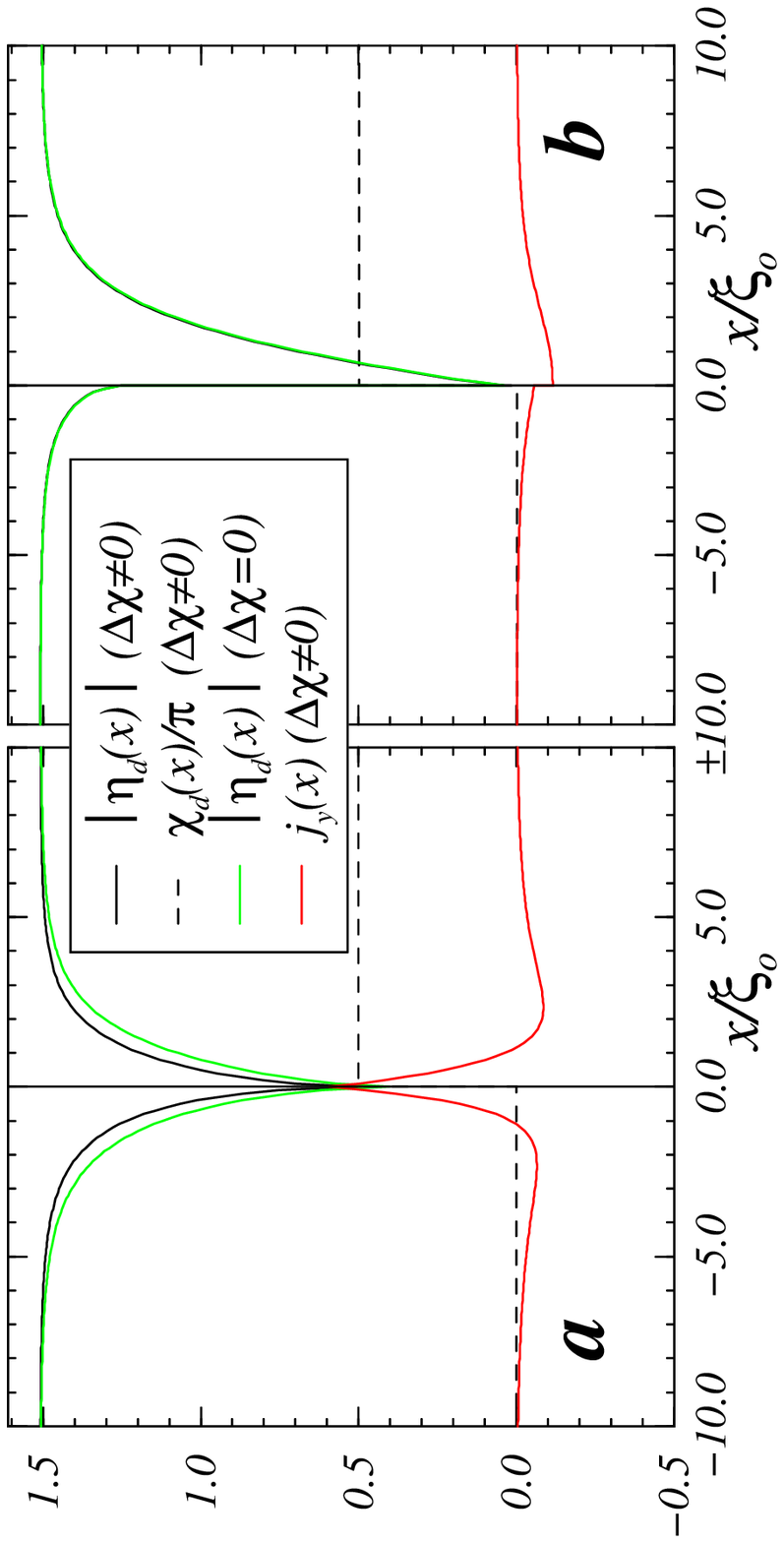} } }

\centerline{\epsfysize=.48\textwidth \rotate[r]
{\epsfbox{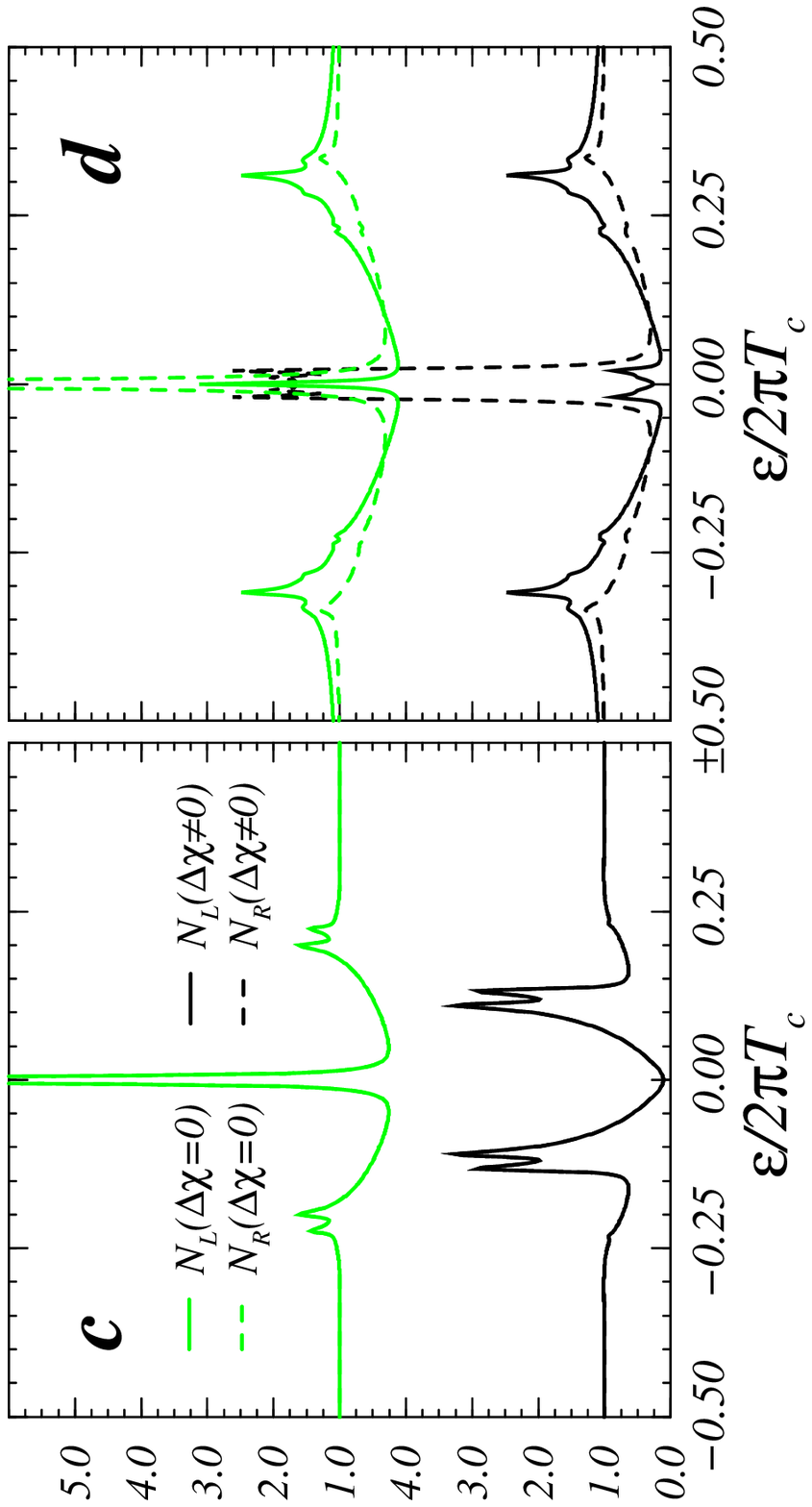} } }
\vskip 0.2 cm
\caption[]
{The magnitude $\vert \eta_d \vert$ and the phase $\chi_d$ of the
order parameter $\eta_d$  for $\alpha_L = 0 $ and
$\alpha_R = \pi/4$ at 
$T = 0.2T_c$ for the states corresponding to $\Delta \chi=0$ and to
the energy minimum with TRSB $\Delta \chi\ne 0$.  
The states with $\Delta \chi \ne 0$ 
have a transverse current density $j_y$ along the boundary.
In the lower panels c and d are 
the corresponding DOS on the two sides of the interface.
${\cal{D}}_0$ is  1.0 in a and c and 0.3 in b and d.
The units are $k_B T_c$ for $\vert \eta_d \vert$
and $2e v_f N_f \vert \eta (\infty )\vert $ for
current densities in all graphs.
$\xi_o \equiv \hbar v_f / 2 \pi T_c$.
}
\label{fig:al0ar4}
\end{figure}

As representative example we consider $\alpha_L = 0$ and $\alpha_R = \pi/4$
at a relatively low temperature, $T=0.2T_c$.
The order parameters are as shown in Fig \ref{fig:al0ar4} 
for both the states with $\Delta \chi=0$ and the ones 
corresponding to energy minima with $\Delta\chi\ne0$.
As can be seen from an examination of Fig. \ref{fig:al0ar4} 
the phase difference $\Delta \chi $ of
the minimum energy state is $ \pi /2$. 
This state is degenerate with its time reversed
partner $- \pi/2$.
The states with $\Delta \chi=0$ are also degenerate 
with the corresponding ones with $\Delta \chi=\pi$
with the same DOS. As claimed the order parameter
of the $\Delta\chi\ne0$ state has a larger amplitude than the
one with zero phase difference for a given transparency.
This difference decreases as ${\cal D}_o$
decreases. At ${\cal D}_o=0.3$ the difference 
in magnitudes is almost undetectable.
The corresponding DOS are shown in
Fig. \ref{fig:al0ar4} for the two different set of states.
The states with $\Delta \chi=0$ have
large DOS near $\varepsilon = 0$. For the state with 
$\Delta\chi\ne0$
these ZEBS are pushed to finite energies, away from $\varepsilon=0$.
These shifts (splits) are largest for ${\cal D}_o = 1$ and
decreases for small ${\cal D}_o$.  As ${\cal D}_o \to 0$ the DOS
becomes independent of $\Delta \chi$.
We also calculated the junction energies for the different states. These
are listed in Table \ref{tab:energies},
which shows explicitly that the $\Delta \chi = \pi/2$ states
have lower energies.
That the TRS-state cannot  
the minimum energy state and that the energy
minimum state is at $\Delta \chi = \pi/2 \ ({\rm mod} \ \pi)$
may actually be expected
from an argument based on symmetry and continuity.\cite{Yip95,Yip93}
It is notable that in the small
transmission limit, ${\cal D}_o=0.3$,
$\Delta \chi = 0$ and $\Delta \chi = \pi/2$
have almost the same free energy. As seen in
Fig. \ref{fig:al0ar4}, the DOS for the two
states are also similar except
some small differences near $\varepsilon=0$. 
At this small ${\cal D}_o$ 
the phase difference $\Delta\chi\ne0$ is inefficient in pushing the 
states that were orginally at $\varepsilon =0$ to finite energies.

%%%%%%%%%%%%%%%%%%%%%%%%%%%%%%%%%%%%%%%%%%%%%%%%%%%%%%%%%%%%%
%%%%%%%%%%%%%%%%%%%%%%%%%%%%%%%%%%%%%%%%%%%%%%%%%%

The DOS recovers to its bulk value 
as one moves away from the interface.
(Fig. \ref{fig:Spatial_DOS}) 
At $x\approx 10\xi_o$ the bulk d-wave DOS is well-recovered showing 
only exponential tails of the structure at the interface.

\begin{minipage}{0.48\textwidth}
%%%%%%%%%%%%%%%%%%%%%%%%%%%%%%%%%%%%%%%%%%%%%%%%%%%%%%%%%%%%%
\begin{figure}[h]
\centerline{ \epsfysize=1.05\textwidth \rotate[r]
{\epsfbox{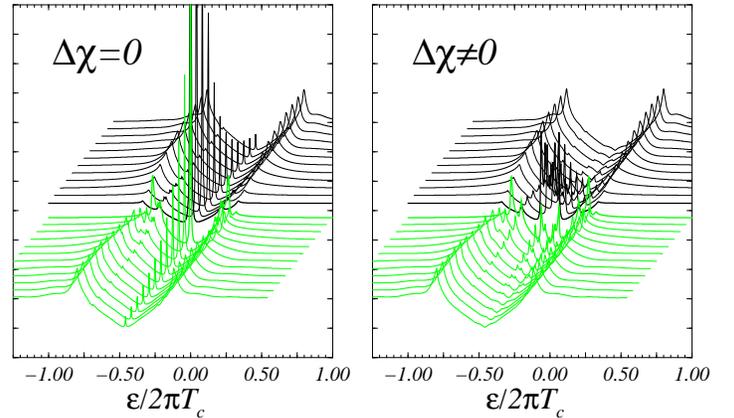} } }
\vskip 0.2 cm
\caption[]
{The spatial dependence of the DOS for an interface with ${\cal D}_o=0.7$. The 
orientation is $(\alpha_L=0,\alpha_R=\frac{\pi}{4})$. The lower (upper)
set of DOS are for 
the left (right) hand side of the interface. The DOS are sampled at
a spacing of $1\xi_o$. The thick lines indicate the
DOS at the  interface location.
}
\label{fig:Spatial_DOS}
\end{figure}

%%%%%%%%%%%%%%%%%%%%%%%%%%%%%%%%%%%%%%%%%%%%%%%%%%%%%%%%%%%%%

\end{minipage}
\vspace*{0.3truecm}

The value of $\Delta \chi$ where the interface
energy is a minimum depends on the orientations
of the crystals, the transmission coefficient
and temperature.  An example is as shown in
Fig \ref{fig:chimin} ({\it c.f.} Refs. 
\cite{Fogelstrom97b,Yip97}). 
The comparison 
between the free energies for the 
orientation $(\alpha_L,\alpha_R) = (-\pi/12,\pi/6)$
is also shown in Table \ref{tab:energies}.

\begin{minipage}{0.48\textwidth}
%%%%%%%%%%%%%%%%%%%%%%%%%%%%%%%%%%%%%%%%%%%%%%%%%%%%%%%%%%%%%
\begin{figure}[h]
\centerline{ \epsfysize=.75\textwidth \rotate[r]
{\epsfbox{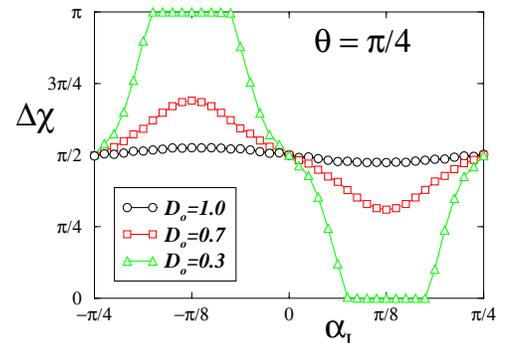} } }
\vskip .2truecm
\caption[]
{
The phase difference $\Delta\chi$ that minimizes the interface 
free energy as a function of $\alpha_L$. The
misorientation angle $\theta$ is kept fixed at $\pi/4$.
The temperature is $0.2 T_c$ 
}
\label{fig:chimin}
\end{figure}
%%%%%%%%%%%%%%%%%%%%%%%%%%%%%%%%%%%%%%%%%%%%%%%%%%
\end{minipage}
\vspace*{0.3truecm}

Recent experiments\cite{Covington97} 
indicate that the oxide
superconductors probably also have an attractive 
s-wave channel with a strength such that
the bare $T_c$ for the s-wave is about
$ 10 \%$ of that of the dominant d-wave.\cite{Fogelstrom97a}
While in the bulk the order parameter is purely $d$ wave,
near the interface both d- and s-wave components can
co-exist.  In this case
\be
\Delta (\hat \vp_f, x) = \eta_d(x) {\sqrt 2} {\rm cos} (2 (\phi - \alpha))
+ \eta_s(x).
\label{mixedstate}
\ee

\begin{figure}[h]
\centerline{ \epsfysize=.48\textwidth \rotate[r]
{\epsfbox{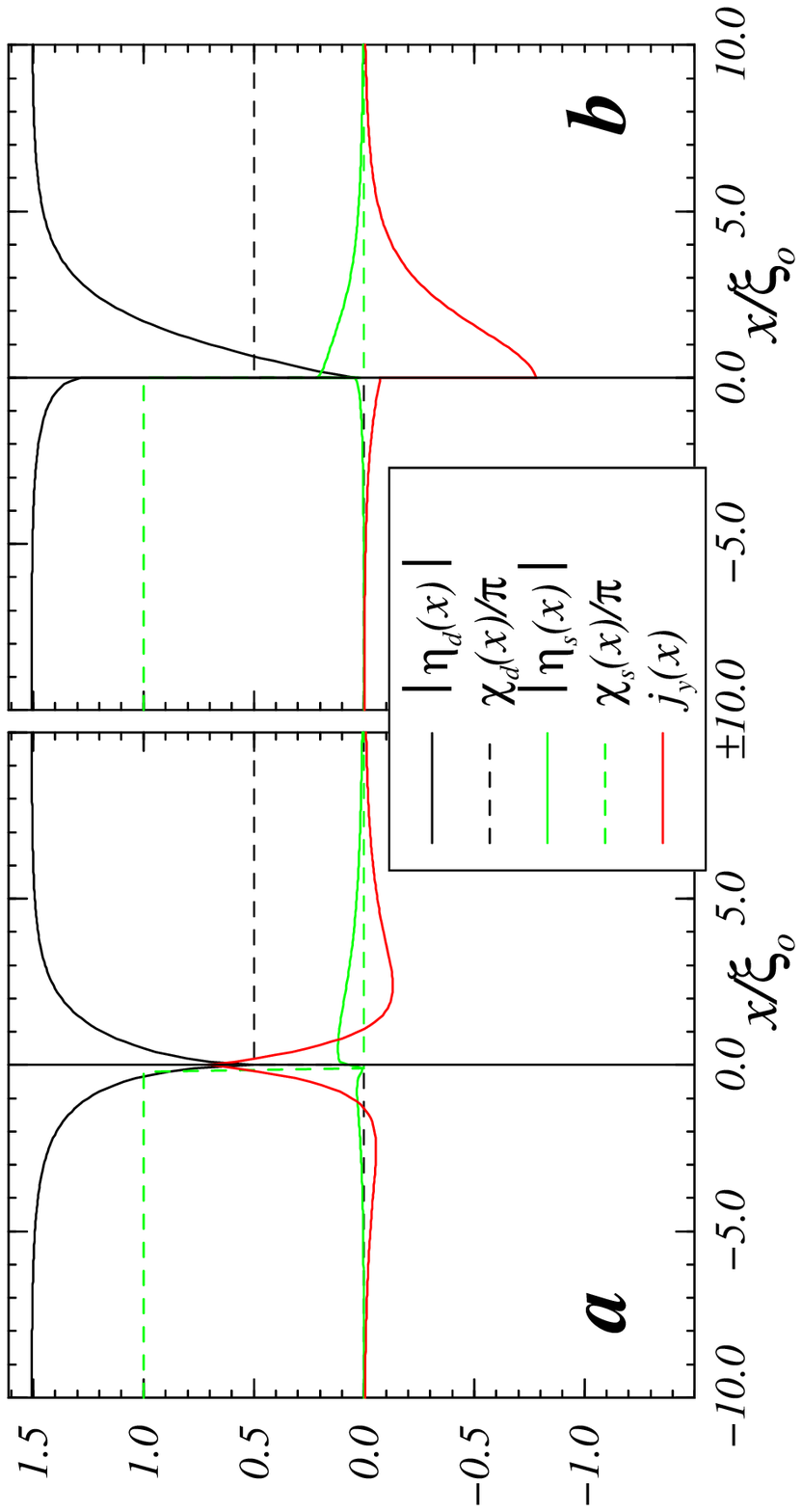} } }

\centerline{ \epsfysize=.48\textwidth \rotate[r]
{\epsfbox{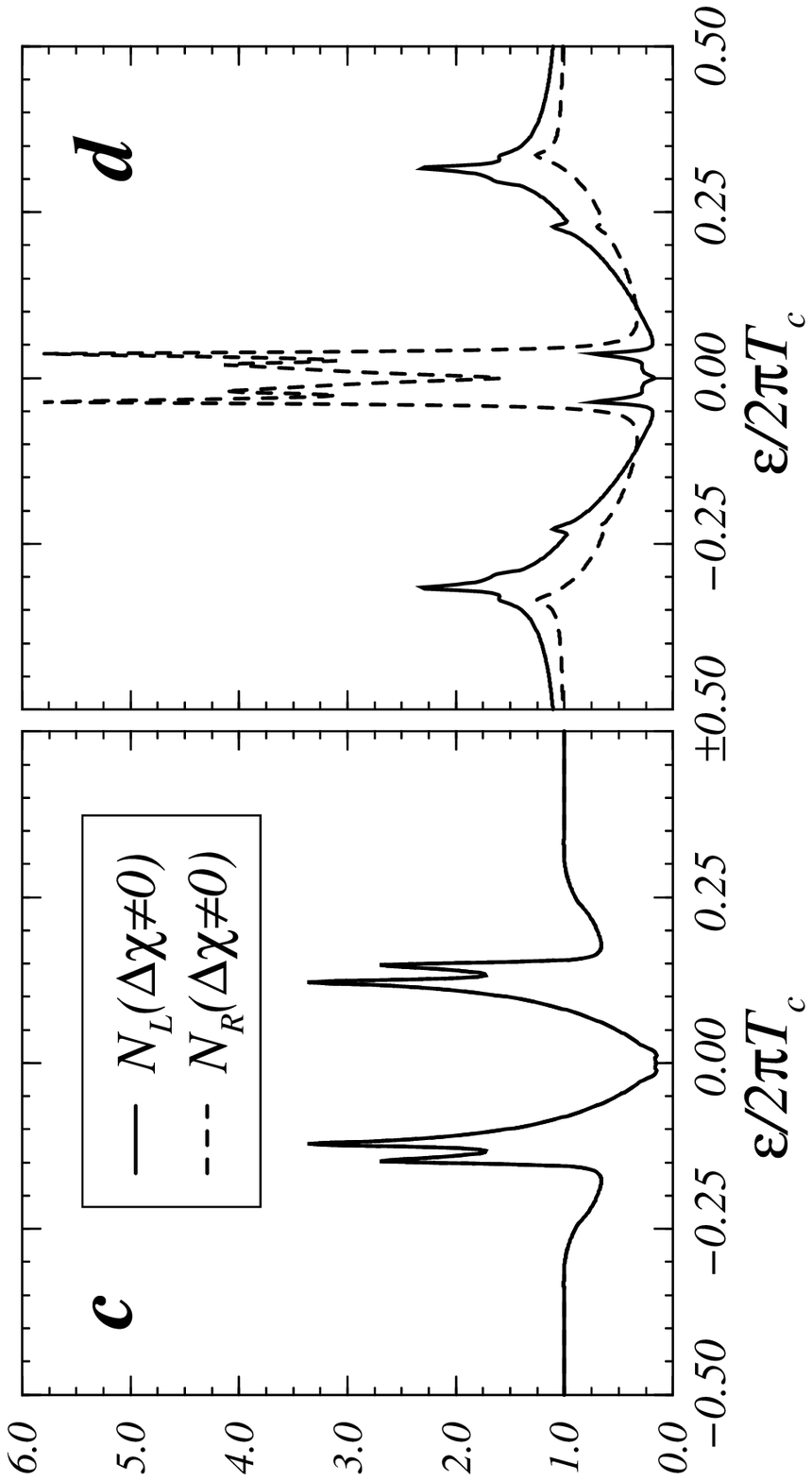} } }

\caption[]{ 
The order parameters $\eta_d$ and $\eta_s$
and the transverse current $j_y$
for $\alpha_L = 0 $ and
$\alpha_R = \pi/4$. $T = 0.2 T_c$ and 
the sub-dominant $T_{c2}=0.1 T_c$. The transparency, ${\cal D}_o$,  of the
boundary is $1.0$ and $0.3$ in panels  a and b.
The corresponding DOS 
at the interfaces are in the lower panels c and d.
}
\label{fig:al0ar4.d+s}
\end{figure}

In Fig \ref{fig:al0ar4.d+s} we have plotted 
the order parameters $\eta_d$, $\eta_s$ 
of the minimum energy
states for $(\alpha_L,\alpha_R) = (0, \pi/4)$ 
and for different transparencies.
For all ${\cal D}_o$ the phase difference between the d-wave 
order parameters on the two sides of the interface
is $\pi/2$ as in the case without the s-wave component of
the order parameter.  In our gauge where the d-wave
order parameter is real for $x \to - \infty$, 
the s-wave component $\eta_s$ is real for all $x$, being
positive for $x > 0$ and negative for $ x < 0$. 
\cite{notesigns} 
The order parameter for $ x > 0$ is in the TRSB
combination $ s + i d$.
For the large transmission ${\cal D}_o = 1$ both the order parameter 
$\eta_d$ and the DOS
are qualitatively equal to the pure $d_{x^2-y^2}$ state shown in Figs. 
\ref{fig:al0ar4}a and c. It is clear that the s-wave
channel does not play an important role. It is rather the tails of the 
off-diagonal parts of the green's function of either side of the interface
that are leaking into the opposite side that give the dominant TRSB.
Hence, for large transmission the main mechanism for
TRSB is the  proximity effect even with a subdominant channel 
of moderate strength present.
As ${\cal D}_o$ is reduced the side with $\alpha=\pi/4$ shows increasing
pair-breaking due to  reflection of  quasiparticles 
by the interface and the TRSB gets more localized to this
right hand 
side. This is also seen in the transverse current density, $j_y(x)$,
which is much larger on this side.  In the small ${\cal D}_o$ limit 
 the proximity effect is gradually shut off and
the presence of the subdominant channel is largely responsible for the TRSB-state.

%%%%%%%%%%%%%%%%%%%%%%%%%%%%%%%%%%%%%%%%%%%%%%%%%%%%%%%%%%%%%
%%%%%%%%%%%%%%%%%%%%%%%%%%%%%%%%%%%%%%%%%%%%%%%%%%

The DOS shown in Figs. \ref{fig:al0ar4}
and \ref{fig:al0ar4.d+s} display considerable 
structure. These results are very different 
from those where the suppression of
the order parameter near the interface is ignored (not shown). 
In particular additional bound states are present
at finite energies.
These bound states are the result of
 Andreev-scattering processes due 
 to amplitude changes in addition to sign
changes in the order parameter.

In conclusion we have investigated time reversal
symmetry breaking at interfaces, in particular
those with high transmission.  We have shown
how this TRSB can be understood from the density
of states and the free energy of the interface.

We thank Juhani Kurkij\"arvi and  Jim Sauls  
for discussions and their comments on the manuscript.
This research was supported by the
the Science and Technology Center for Superconductivity, 
grant no. NSF 91-20000,  
the Academy of Finland 
research grant No. 4385 (SY),
and the {\AA}bo Akademi. 
MF also acknowledges partial support from SF{\AA}AF
 and Magnus Ehrnrooths Stiftelse.


\begin{thebibliography}{10}

\bibitem{Covington97} M.~Covington {\it et~al.},
        {\it Phys. Rev. Lett.} {\bf 79}, 277 (1997)

\bibitem{Fogelstrom97a} M.~Fogelstr\"om, D.~Rainer and J.~A.~Sauls,
        {\it Phys. Rev. Lett.} {\bf 79}, 281 (1997); M.~Fogelstr\"om,
        M.~Palumbo, L.~Buchholtz , D.~Rainer and  J.~A.~Sauls, preprint

\bibitem{Kirtley96}
        J.~Kirtley {\it et~al.} {\it Phys. Rev. Lett.} {\bf 76} 1336 (1996)
          
\bibitem{Sigrist95}
        M.~Sigrist, D.~B.~Bailey and R.~B.~Laughlin, 
        {\it Phys. Rev. Lett.} {\bf 74} 3249 (1995)

\bibitem{Yip95} 
        S.-K.~Yip, {\it Phys. Rev. B} {\bf 52}, 3087 (1995)

\bibitem{Sigrist96}
        M.~Sigrist, K.~Kuboki, B.~Kuklov, D.~B.~Bailey  
        and R.~B.~Laughlin, {\it Czech. J. of Physics}, {\bf 46}, 3159 (1996)

\bibitem{Bailey97}
        D.~B.~Bailey, M.~Sigrist and R.~B.~Laughlin ,
        {\it Phys. Rev. B} {\bf 55}, 15239 (1997)

\bibitem{Buchholtz95}
          L.~Buchholtz, M.~Palumbo, D.~Rainer and J.~A.~Sauls,
          {\it J. Low Temp. Phys.} {\bf 101}, 1079 (1995); 
          {\it ibid.} {\bf 101}, 1099 (1995)

\bibitem{Matsumoto95}
        M.~Matsumoto and H.~Shiba, 
       {\it J. Phys. Soc. Japan} {\bf 64}, 1703 (1995);
       {\it ibid } {\bf 64} 3384 (1995); {\it ibid} {\bf 64} 4847 (1995)

\bibitem{Barash95}
        Yu~S.~Barash, A.~V.~Galaktionov and A.~D.~Zaikin
        {\it Phys. Rev. B} {\bf 52}, 665 (1995)

\bibitem{Fogelstrom97b} M.~Fogelstr\"om, S.-K.~Yip and J.~Kurkij\"arvi,
        {\it Physica C}, to appear (1998), cond-mat/9709120

\bibitem{Yip97} 
          S.-K.~Yip, {\it J. Low Temp. Phys.}, {\bf 109}, 547 (1997)

\bibitem{proof} M.~Fogelstr\"om and  S.-K.~Yip, unpublished

\bibitem{notebruder}  Recently, Belzig {\it et~al.} (cond-mat/9801107)
have considered
TRSB near a twin boundary from a similar point of view.

\bibitem{Serene83}
          J.~W.~Serene and D.~Rainer {\it Phys. Rep.} {\bf 101} 221 (1983)

\bibitem{Thuneberg84}
        E.~V.~Thuneberg, J.~Kurkij\"arvi and D.~Rainer,
       {\it Phys. Rev. B} {\bf 29}, 3913 (1984)

\bibitem{Yip93} 
        S.-K.~Yip, {\it J. Low Temp. Phys.} {\bf 91}, 203 (1993)

\bibitem{notesigns}  The sign of the order parameter $\eta_s$
can be understood by considering the leakage of
the off-diagonal part of the green's function
from one side to the other.

\end{thebibliography}
\end{document}